# Two-Dimensional Multiferroic Semiconductors with Coexisting Ferroelectricity and Ferromagnetism

Jingshan Qi[1,a)], Hua Wang[2], Xiaofang Chen[1], and Xiaofeng Qian[2,a)]

[1]*School of Physics and Electronic Engineering, Jiangsu Normal University, Xuzhou 221116, P.R. China*

[2]*Department of Materials Science and Engineering, Texas A&M University, College Station, Texas 77843, USA*

Abstract

Low-dimensional multiferroicity, though highly scarce in nature, has attracted great attention due to both fundamental and technological interests. Using first-principles density functional theory, we show that ferromagnetism and ferroelectricity can coexist in monolayer transition metal phosphorus chalcogenides (TMPCs) - $CuMP_2X_6$ (M=Cr, V; X=S, Se). These van der Waals layered materials represent a class of 2D multiferroic semiconductors that simultaneously possess ferroelectric and ferromagnetic orders. In these monolayer materials, Cu atoms spontaneously move away from the center atomic plane, giving rise to nontrivial electric dipole moment along the plane normal. In addition, their ferromagnetism originates from indirect exchange interaction between Cr/V atoms, while their out-of-plane ferroelectricity suggests the possibility of controlling electric polarization by external vertical electric field. Monolayer semiconducting TMPCs thus provide a solid-state 2D materials platform for realizing 2D nanoscale switches and memory devices patterned with top and bottom electrodes.

[a)] Authors to whom correspondence should be addressed. Electronic addresses: qijingshan@jsnu.edu.cn and feng@tamu.edu

Multiferroics are a special class of materials that simultaneously possess two or more primary ferroic orders among ferroelasticity, ferroelectricity (FE), ferromagnetism (FM), and ferrotoroidicity. The last three requires breaking spatial-inversion symmetry, time-reversal symmetry, and both, respectively. Among the four ferroic orders, FM and FE are crucial for technological applications such as magnetic memory[1] and ferroelectric memory.[2,3] Moreover, the materials with coexisting FM and FE is called magnetoelectric multiferroics.[4,5] Magnetism usually comes from the ordered spins of electrons in the partially filled *d*/*f* orbitals of transition metals, while FE often results from residual polarization due to stable off-centered ion with empty *d*/*f* orbitals in transition metal. It is, therefore, not surprising that many ferromagnets tend to be metallic while such metallic nature will screen out electric polarization. The above mutually exclusive requirements on the electronic occupation of the outer valence shells makes it particularly challenging to host FE and FM within the same material. Consequently, only a limited number of multiferroic materials have been discovered to date.[4,5]

Low-dimensional multiferroicity is even more attractive owing to their potential applications in miniaturized devices. However, maintaining the stability of FE and FM at room temperature is another nontrivial issue for ultrathin films. In particular, the depolarization field is usually enhanced in thin films below a few nanometers (e.g. about 12 Å and 24 Å for conventional FE materials $BaTiO_3$ or $PbTiO_3$, respectively).[6,7] This depolarization field will compete with intrinsic ferroelectricity. Very recently several groups report the discoveries of two-dimensional (2D) FM,[8,9] 2D FE,[10-19] 2D ferroelastic,[20,21] 2D ferroelastic-ferromagnetic[22] and 2D ferroelastic-ferroelectric[23,24] materials. In addition, large second harmonic generation (SHG)[25,26] and photostriction[27] were also discovered in 2D ferroics. Remarkably, giant ferroelectricity and ferroelasticity in monolayer group IV monochalcogenides were found to be strongly coupled with SHG,[32] making SHG a unique tool for distinguishing different ferroelectric and ferroelastic orders. These findings suggest a possibility of achieving 2D multiferroicity with coexisting FM and FE orders which are highly desired for both fundamental science and technological applications.

Using first-principles theory, herein we predict that the monolayer transition metal phosphorus chalcogenides TMPCs - $CuMP_2X_6$ (M=Cr, V; X=S, Se) represent a class of 2D multiferroic materials that simultaneously possess FE and FM orders. Their FE order comes from the out-of-plane electric dipole moment due to energetically favorable spontaneous displacement of Cu atoms away from the center atomic plane. It is worth to note that, although antiferroelectricity (AFE) states are more stable energetically than FE and PE states for four compounds, the transition energy barrier between FE and AFE is large enough to prevent spontaneous transition from the FE to AFE state at room temperature in monolayer $CuCrP_2S_6$ and $CuVP_2S_6$. Their intermediate barrier indicates that FE states should be observed in experiment under ambient conditions. Furthermore, the magnetism originates from Cr or V atoms. Our first-principles calculations suggest that the FM state is more stable than the antiferromagnetic (AFM) and non-magnetic (NM) states. The large magnetocrystalline anisotropy energy (MAE) enables the presence of long-range FM order in these 2D materials. We would like to emphasize that the intrinsic *out-of-plane* electric polarization in TMPCs is different from the *in-plane* polarization recently found in monolayer group IV monochalcogenides[23, 24] and atomic-thick SnTe.[10] Finally, all monolayer TMPCs are semiconducting crystals, which is beneficial to the coexistence of FE and FM states.

Electronic structure calculations of monolayer TMPCs with different lattice constant were carried out using first-principles density-functional theory (DFT)[28, 29] as implemented in the Vienna Ab initio Simulation Package (VASP).[30] We use the exchange-correlation functional in the Perdew-Burke-Ernzerhof (GGA) form[31] within the generalized gradient approximation and adopt the projector-augmented wave method.[32] Additionally, we apply an energy cutoff of 400 eV for the plane-wave basis and a 12×12×1 Monkhorst-Pack grid for k-point sampling.[33] A vacuum slab of 20 Å is included to minimize the image interaction from periodic boundary condition. Atomic coordinates were optimized with maximum residual force of 0.01 eV/Å. To properly take into account the interlayer van der Waals interaction, we employed the Grimme's semi-empirical correction method.[34] In addition, we apply Hubbard U corrections to account for electronic correlation in 3$d$ transition metals.[35] Below we will show the results obtained from GGA+U calculations with a moderate effective value, $U_{eff}$=U-J=3 eV. The results calculated with different $U_{eff}$ values can be found in the Supplementary Material (SM).

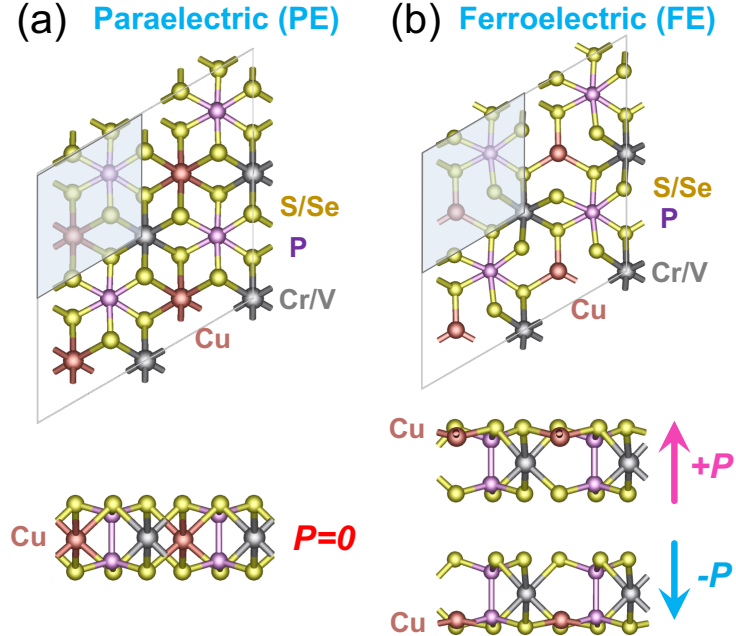

**FIG. 1**. Crystal structure of monolayer TMPCs - $CuMP_2X_6$ (M=Cr, V; X=S, Se). (a) Paraelectric (PE) phase. (b) Ferroelectric (FE) phase.

TMPCs - $CuMP_2X_6$ (M=Cr, V; X=S, Se) are van der Waals layered compounds. They were synthesized more than three decades ago.[16, 36-40] Monolayer TMPC consists of a chalcogenide (S/Se) framework with the octahedral sites filled by Cu, Cr/V and P–P in a triangular pattern, as illustrated in Fig. 1. Paraelectric (PE) and FE phases are shown in Figs. 1(a) and 1(b), respectively. In the PE phase, Cu and Cr/V atoms are all located in the middle plane. In the FE phase, Cu atoms are displaced along the plane normal, accompanied with spontaneous out-of-plane polarization. The FE polarization in these materials is similar to the recently discovered ferroelectric layered compound $CuInP_2S_6$,[16] however the latter does not possess magnetism.

Variations of total energy with respect to different lattice constant are presented in Fig. S1 for four monolayer TMPCs - $CuMP_2X_6$ (M=Cr, V; X=S, Se). It can be seen that two stable/metastable phases of PE and FE exist. To quantify the out-of-plane ferroelectricity we calculate the electric dipole moment along the plane normal. It is 0.79, 0.67, 0.78 and 0.65 $pC/m$ for monolayer $CuCrP_2S_6$, $CuCrP_2Se_6$, $CuVP_2S_6$, and $CuVP_2Se_6$, respectively. In addition, it should be noted that the Hubbard U parameter ($U_{eff}$) used in the calculations does have some influence on the relative energy difference of the FE and PE phases. However two FE and PE phases always exist for a wide range of $U_{eff}$ from 0 to 3 eV, and the FE phase of monolayer $CuCrP_2S_6$ is always more stable than the PE phase (see Fig. S1 in SM).

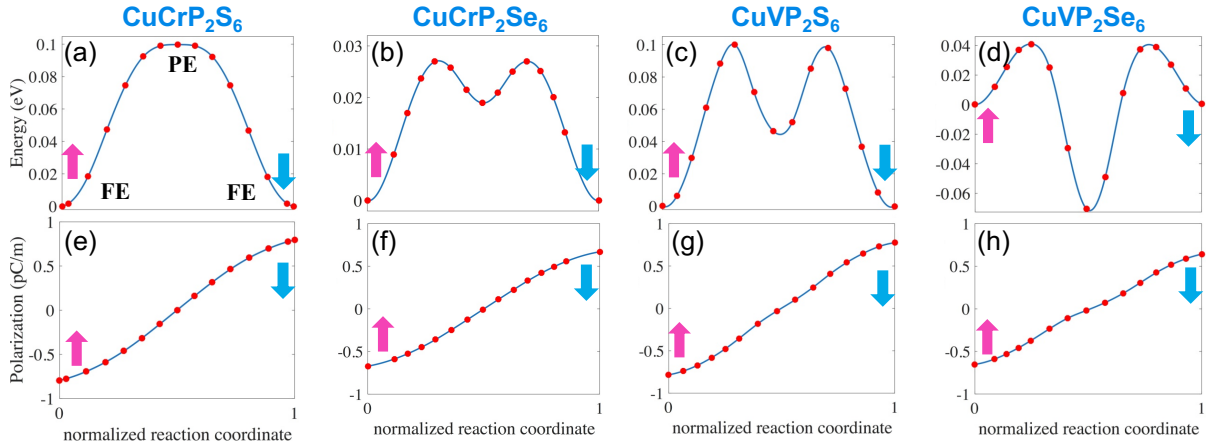

**FIG. 2**. (a-d) Minimum energy pathway of FE-PE-FE transition in monolayer TMPCs - $CuMP_2X_6$. Energy value here is for a single formula unit. (e-h) Evolution of total electric polarization along the FE-PE-FE transition pathway.

To understand the transition between the FE and PE phase, we further investigate the minimum energy pathway and the corresponding polarization from FE to PE phase. This is achieved by using the climbing image nudged elastic band (CI-NEB) method[41] with total energy and interatomic forces calculated by DFT. The corresponding results for four TMPC monolayers are shown in Fig. 2. We can see that the minimum energy barriers from the FE to PE state are about 100, 27, 100, and 41 meV/formula unit (f.u.) for $CuCrP_2S_6$, $CuCrP_2Se_6$, $CuVP_2S_6$, and $CuVP_2Se_6$, respectively. These potential barriers are reasonably large (*e.g.* 100 meV for $CuCrP_2S_6$ and $CuVP_2S_6$) and make the metastable states dynamically stable. For comparison, the barrier height for conventional FE material $PbTiO_3$ is about 75 meV.[42] So it is possible to separate the FE and PE phase in experiment. In addition, an appropriate intermediate barrier is necessary for facile switching of ferroelectricity. Therefore, these small transition barriers are suitable for 2D ferroelectric nanoscale switch and memory applications. It is important to notice that both FE and PE phases of $CuCrP_2Se_6$, $CuVP_2S_6$ and $CuVP_2Se_6$ are either globally stable or metastable with two saddle points, as evidenced in Fig. 2. Although first-principles calculations here provided potential energy barrier, phase transition temperature is a more complicated physical quantity, not only related to kinetic barrier but also depending on appropriate theoretical model. For example, Barraza-Lopez et al.[19] recently found that first-principles molecular dynamics simulations show qualitative agreement with four-state 2D clock model for group IV monochalcogenide monolayers, indicating an order-disorder phase transition that may not be described by Landau theory in two dimension. Therefore, specific phase transition temperature in the four TMPCs here requires a separate investigation of the interplay among the dimensionality, symmetry, boundary conditions, and specific structures, which is beyond the present scope.

We further studied AFE states in a 2×1 supercell by moving one Cu atom up and the other down away from the center atomic plane. Figure S2 in supplementary material shows the calculated migration energy barrier from a FE to AFE state using the CI-NEB method. AFE states are found to be more stable than FE and PE states for four compounds. However, the minimum energy barriers from FE to AFE state are 34 meV/f.u. and 40 meV/f.u. for $CuCrP_2S_6$ and $CuVP_2S_6$, respectively. These kinetic barriers are large enough to prevent spontaneous transition from FE to AFE state at room temperature. For $CuCrP_2Se_6$ and $CuVP_2Se_6$, the minimum energy barriers from FE to AFE state are 3.5 meV/f.u. and 11 meV/f.u., respectively, indicating that AFE state can be observed in experiment. Therefore, $CuCrP_2S_6$ and $CuVP_2S_6$ are more promising for practical ferroelectric applications.

Figure 3(a) shows spin density of FM state in monolayer $CuCrP_2S_6$. It demonstrates that the magnetic moment is mainly located at Cr/V atoms. Each Cr (V) atom has about 3.0 $\mu_B$ (2.0 $\mu_B$) moment. Magnetic orders in their monolayers were studied by using a 2×2 supercell. NM, FM and several AFM states were considered in total energy calculations. The FM state was found to be always the most stable one, regardless of FE or PE phase. The FM ordering originates from the indirect exchange interaction between Cr/V atoms mediated by Cu and S/Se. Energy difference between FM and AFM is about 14 meV per Cr atom for both $CuCrP_2S_6$ and $CuCrP_2Se_6$, and is about 1.5meV per V atom for both $CuVP_2S_6$ and $CuVP_2Se_6$. The energy of NM is much higher than those of FM and AFM by about 3.2 eV/f.u. for $CuCrP_2S_6$ and $CuCrP_2Se_6$ and about 1.7 eV/f.u. for $CuVP_2S_6$ and $CuVP_2Se_6$. According to the Mermin-Wagner theorem,[43] FM order is prohibited within 2D isotropic Heisenberg model with continuous SU(2) or SO(3) symmetry. However, finite magnetic anisotropy will help stabilize long-range magnetic order. To determine the MAE, we calculate the total energy as a function of magnetization direction. Figure 3(b) presents the angular dependence of the MAE in monolayer $CuCrP_2S_6$, while the result for $CuCrP_2Se_6$ is shown in Fig. S3. These data demonstrate that both monolayer $CuCrP_2Se_6$ and $CuCrP_2S_6$ possess high magnetic anisotropy. The energy difference between in-plane and out-of-plane magnetization, $\Delta E \equiv E_{x/y} - E_z$, is -28 $\mu$eV and +242 $\mu$eV for $CuCrP_2S_6$ and $CuCrP_2Se_6$, respectively. This indicates that the easy axis of monolayer $CuCrP_2Se_6$ is on the 2D plane, while the easy axis of monolayer $CuCrP_2S_6$ is along the plane normal. These MAE values are much higher than those of nickel, iron, and cobalt, three well-known room-temperature magnetic element crystals with the MAE on the order of 1 $\mu$eV.[44] Our results therefore confirm that the FM and FE phases can coexist in these single-phase 2D materials. Very recently, 2D magnetic semiconductors, $CrGeTe_3$[8] and $CrI_3$[9] have been shown to maintain their ferromagnetism at single atomic-layer level, although ferroelectricity is absent in these 2D semiconductors.

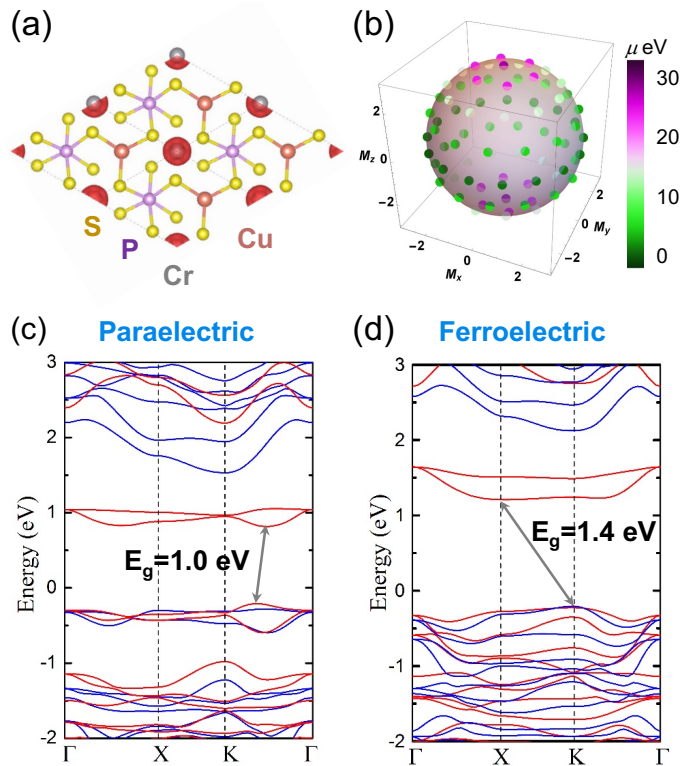

**FIG. 3**. Electronic structure of monolayer $CuCrP_2S_6$. (a) Spin density plot of the FE phase. (b) Magnetic anisotropy of monolayer $CuCrP_2S_6$ as function of the magnetization direction. (c-d) Electronic band structure of FE and PE phase, respectively.

The FE-PE phase transition of these monolayer compounds discussed above will have more profound impact if there exists direct coupling between the FE/PE phase and their electronic/optical properties. Figure 3(c) and 3(d) show the band structures of the PE and FE phase for monolayer $CuCrP_2S_6$, respectively, where the red and blue colors indicate spin-majority and spin-minority electron, respectively. Electronic band structure for $CuCrP_2Se_6$, $CuVP_2S_6$, and $CuVP_2Se_6$ can be found in Fig. S4. The results demonstrate that all four monolayer materials are intrinsic 2D semiconducting materials with indirect band gap ranging from 0.9 eV to 1.4 eV. Although the band structure profiles are similar, the band gaps of the FE and PE phases are clearly different. The calculated band gaps of PE (FE) phase are about 1.0 (1.4), 0.9 (1.25), 1.0 (1.3), and 0.9(1.0) eV for $CuCrP_2S_6$, $CuCrP_2Se_6$, $CuVP_2S_6$, and $CuVP_2Se_6$, respectively. The intrinsic coupling between FE/PE phase and the electronic band gap implies that these materials may be used for ferroelectric switching and memory device applications.

The above findings point to the possibility of several compelling device concepts. First, by applying external vertical electric field one can control the FE-PE phase transition. Then taking advantage of different band gap in the FE and PE phases the FE/PE states can be detected by measuring electric current under a small bias, thereby realizing 2D ferroelectric memory as shown in Fig. 4(a). Second, we can also control the two FE states with opposite polarization direction by reversing external vertical electric field. The corresponding states can be detected or "read" by measuring the surface photocurrent or surface photovoltage under photo-illumination. This is because the opposite internal electric field from the opposite FE polarization will drive photoexcited charge carriers (*i.e.* electron and hole) to drift in opposite direction, as presented in Fig. 4(b). In addition, these monolayer materials may be useful for realizing 2D ferroelectric photovoltaics, as their band gap (0.9 - 1.4 eV) falls within the visible range and the internal electric field in ferroelectric monolayers may enhance the separation of photo-excited charge carriers.

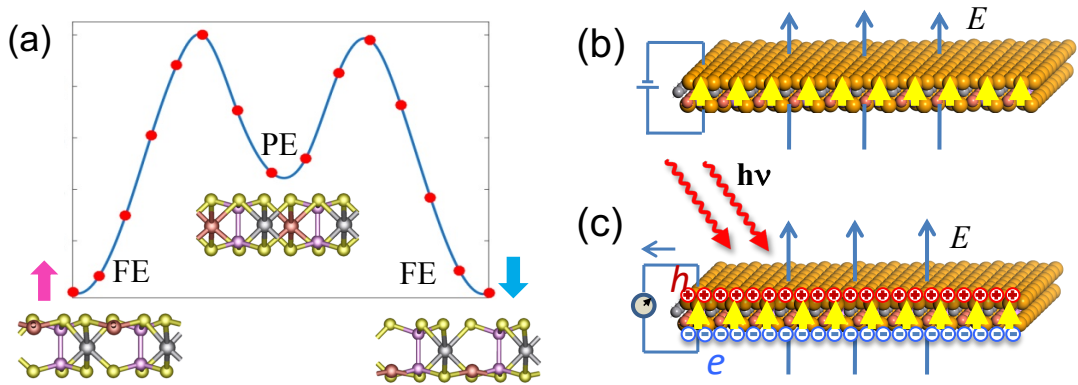

**FIG. 4**. 2D TMPCs - $CuMP_2X_6$ (M=Cr, V; X=S, Se) based ferroelectric and optoelectronic device concepts. (a) Energy landscape of FE-PE-FE transition in 2D TMPCs. (b) Schematic of 2D ferroelectric memory. (c) Schematic of 2D ferroelectric optoelectronic device.

In summary, in this Letter we predict that monolayer TMPCs - $CuMP_2X_6$ (M=Cr, V; X=S, Se) represent a class of 2D multiferroic materials with the coexistence of FE and FM orders. Spontaneous polarization of the FE phase originates from the low energy configuration where Cu atoms are displaced away from the center plane, forming an out-of-plane ferroelectricity. Although AFE states are energetically more stable than FE and PE states, the energy barrier in monolayer $CuCrP_2S_6$ and $CuVP_2S_6$ is large enough to prevent spontaneous transition from FE to AFE state at room temperature. This indicates that the FE state can be observed in experiment. Furthermore, the FM order comes from indirect exchange interaction between Cr/V atoms which is more stable than AFM order. The large magnetic anisotropy suggests the possibility of long-range FM order in monolayer TMPCs. Our results also show that these compounds are all semiconducting with the band gap ranging from 0.9 eV to 1.4 eV, making them suitable for 2D ferroelectric photovoltaic applications. Finally, the FE and PE states may be further controlled by external vertical electric field, which may open avenues for 2D multiferroics-based nanoscale memories and switches.

**Supplementary Material:** details on the total energy of FE and PE phases of monolayer TMPCs as function of lattice constant, minimum energy pathway for FE-AFE transition, MAE as function of magnetization direction, and electronic band structure of four TMPCs.

J.Q. acknowledges the financial support from the National Natural Science Foundation of China (Projects No. 11674132) and PAPD. H.W. and X.Q. acknowledge the start-up funds from Texas A&M University and the support from Texas A&M Energy Institute. Portions of this research were conducted with the advanced computing resources provided by Texas A&M High Performance Research Computing.

# References

1   J. Åkerman, Science **308**, 508-510 (2005).
2   M. Dawber, K. M. Rabe and J. F. Scott, Rev. Mod. Phys. **77**, 1083-1130 (2005).
3   L. W. Martin and A. M. Rappe, Nat. Rev. Mater. **2**, 16087 (2016).
4   R. Ramesh and N. A. Spaldin, Nat. Mater. **6**, 21-29 (2007).
5   W. Eerenstein, N. D. Mathur and J. F. Scott, Nature **442**, 759-765 (2006).
6   D. D. Fong, G. B. Stephenson, S. K. Streiffer, J. A. Eastman, O. Auciello, P. H. Fuoss and C. Thompson, Science **304**, 1650-1653 (2004).
7   J. Junquera and P. Ghosez, Nature **422**, 506-509 (2003).
8   C. Gong, L. Li, Z. Li, H. Ji, A. Stern, Y. Xia, T. Cao, W. Bao, C. Wang, Y. Wang, Z. Q. Qiu, R. J. Cava, S. G. Louie, J. Xia and X. Zhang, Nature **546**, 265-269 (2017).
9   B. Huang, G. Clark, E. Navarro-Moratalla, D. R. Klein, R. Cheng, K. L. Seyler, D. Zhong, E. Schmidgall, M. A. McGuire, D. H. Cobden, W. Yao, D. Xiao, P. Jarillo-Herrero and X. Xu, Nature **546**, 270-273 (2017).
10  K. Chang, J. Liu, H. Lin, N. Wang, K. Zhao, A. Zhang, F. Jin, Y. Zhong, X. Hu, W. Duan, Q. Zhang, L. Fu, Q.-K. Xue, X. Chen and S.-H. Ji, Science **353**, 274-278 (2016).
11  M. Mehboudi, B. M. Fregoso, Y. Yang, W. Zhu, A. van der Zande, J. Ferrer, L. Bellaiche, P. Kumar and S. Barraza-Lopez, Phys. Rev. Lett. **117**, 246802 (2016).
12  R. Fei, W. Kang and L. Yang, Phys. Rev. Lett. **117**, 097601 (2016).
13  S. N. Shirodkar and U. V. Waghmare, Phys. Rev. Lett. **112**, 157601 (2014).
14  M. Wu, S. Dong, K. Yao, J. Liu and X. C. Zeng, Nano Lett. **16**, 7309-7315 (2016).
15  E. Bruyer, D. Di Sante, P. Barone, A. Stroppa, M.-H. Whangbo and S. Picozzi, Phys. Rev. B **94**, 195402 (2016).
16  F. Liu, L. You, K. L. Seyler, X. Li, P. Yu, J. Lin, X. Wang, J. Zhou, H. Wang, H. He, S. T. Pantelides, W. Zhou, P. Sharma, X. Xu, P. M. Ajayan, J. Wang and Z. Liu, Nat. Commun. **7**, 12357 (2016).
17  W. Wan, C. Liu, W. Xiao and Y. Yao, Appl. Phys. Lett. **111**, 132904 (2017).

18  W. Ding, J. Zhu, Z. Wang, Y. Gao, D. Xiao, Y. Gu, Z. Zhang and W. Zhu, Nat. Commun. **8**, 14956 (2017).
19  S. Barraza-Lopez, T. P. Kaloni, S. P. Poudel and P. Kumar, Phys. Rev. B **97**, 024110 (2018).
20  W. Li and J. Li, Nat. Commun. **7**, 10843 (2016).
21  M. Mehboudi, A. M. Dorio, W. Zhu, A. van der Zande, H. O. Churchill, A. A. Pacheco-Sanjuan, E. O. Harriss, P. Kumar and S. Barraza-Lopez, Nano Lett. **16**, 1704-1712 (2016).
22  L. Seixas, A. S. Rodin, A. Carvalho and A. H. Castro Neto, Phys. Rev. Lett. **116**, 206803 (2016).
23  M. Wu and X. C. Zeng, Nano Lett. **16**, 3236-3241 (2016).
24  H. Wang and X. Qian, 2D Mater. **4**, 015042 (2017).
25  H. Wang and X. Qian, Nano Lett. **17**, 5027-5034 (2017).
26  P. Suman Raj and M. F. Benjamin, J. Phys.: Condens. Matter **29**, 43LT01 (2017).
27  R. Haleoot, C. Paillard, T. P. Kaloni, M. Mehboudi, B. Xu, L. Bellaiche and S. Barraza-Lopez, Phys. Rev. Lett. **118**, 227401 (2017).
28  P. Hohenberg and W. Kohn, Phys. Rev. **136**, B864-B871 (1964).
29  W. Kohn and L. J. Sham, Phys. Rev. **140**, A1133-A1138 (1965).
30  G. Kresse and J. Furthmüller, Phys. Rev. B **54**, 11169-11186 (1996).
31  J. P. Perdew, K. Burke and M. Ernzerhof, Phys. Rev. Lett. **77**, 3865-3868 (1996).
32  P. E. Blöchl, Phys. Rev. B **50**, 17953-17979 (1994).
33  H. J. Monkhorst and J. D. Pack, Phys. Rev. B **13**, 5188-5192 (1976).
34  S. Grimme, J. Comput. Chem. **27**, 1787-1799 (2006).
35  S. L. Dudarev, G. A. Botton, S. Y. Savrasov, C. J. Humphreys and A. P. Sutton, Phys. Rev. B **57**, 1505 (1998).
36  X. Bourdon, V. Maisonneuve, V. B. Cajipe, C. Payen and J. E. Fischer, J. Alloys Compd. **283**, 122-127 (1999).
37  R. Brec, Solid State Ion. **22**, 3-30 (1986).
38  V. Maisonneuve, M. Evain, C. Payen, V. B. Cajipe and P. Molinié, J. Alloys Compd. **218**, 157-164 (1995).
39  M. A. Susner, M. Chyasnavichyus, M. A. McGuire, P. Ganesh and P. Maksymovych, Adv. Mater. **29**, 1602852 (2017).
40  A. Belianinov, Q. He, A. Dziaugys, P. Maksymovych, E. Eliseev, A. Borisevich, A. Morozovska, J. Banys, Y. Vysochanskii and S. V. Kalinin, Nano Lett. **15**, 3808-3814 (2015).
41  G. Henkelman, B. P. Uberuaga and H. Jónsson, J. Chem. Phys. **113**, 9901-9904 (2000).
42  S. P. Beckman, X. Wang, K. M. Rabe and D. Vanderbilt, Phys. Rev. B **79**, 144124 (2009).
43  N. D. Mermin and H. Wagner, Phys. Rev. Lett. **17**, 1133-1136 (1966).
44  S. V. Halilov, A. Y. Perlov, P. M. Oppeneer, A. N. Yaresko and V. N. Antonov, Phys. Rev. B **57**, 9557-9560 (1998).